\documentclass[final,numberedheadings]{aipproc}

\usepackage{amsmath,amssymb}
\layoutstyle{6x9}

\newcommand{\be}{\begin{equation}}
\newcommand{\ee}{\end{equation}}
\newcommand{\bea}{\begin{eqnarray}}
\newcommand{\eea}{\end{eqnarray}}

\newcommand{\nn}{\nonumber}

\newcommand{\uno}{1 \!\!\! 1}

\newcommand{\R}{{\kern+.25em\sf{R}\kern-.78em\sf{I} \kern+.78em\kern-.25em}}
\newcommand{\RR}{{\kern+.25em\sf{R}\kern-.6em\sf{I} \kern+.6em\kern-.25em}}
\newcommand{\N}{{\kern+.25em\sf{N}\kern-.78em\sf{I} \kern+.78em\kern-.25em}}
\newcommand{\C}{{\kern+.25em\sf{C}\kern-.45em\sf{I} \kern+.45em\kern-.25em}}

\def\lsi{\raise0.3ex\hbox{$<$\kern-0.75em\raise-1.1ex\hbox{$\sim$}}}
\def\gsi{\raise0.3ex\hbox{$>$\kern-0.75em\raise-1.1ex\hbox{$\sim$}}}

\newcommand{\lsim}{\mathop{\lsi}}
\newcommand{\gsim}{\mathop{\gsi}}

\def\backder{\raise1.4ex\hbox{$\leftarrow$\kern-0.75em\raise-1.4ex\hbox{$\partial$}}}
\def\forder{\raise1.4ex\hbox{$\rightarrow$\kern-0.75em\raise-1.4ex\hbox{$\partial$}}}
\newcommand{\backderi}{\mathop{\backder}}
\newcommand{\forderi}{\mathop{\forder}}

\begin{document}

\title{The Photon Dispersion as an \\ Indicator for New Physics ?}

\classification{11.10.Lm, 11.10.Nx, 11.15.Ha, 11.30.Cp, 11.55.Fv, 14.70.Bh}
%http://www.aip.org/pacs/index.html
%11.10.Lm 	Nonlinear or nonlocal theories and models
%11.10.Nx 	Noncommutative field theory
%11.15.Ha 	Lattice gauge theory
%11.30.Cp 	Lorentz and Poincaré invariance
%11.55.Fv 	Dispersion relations
%12.20.Fv 	QED: Experimental tests 
%14.70.Bh 	Photons

\keywords{Photon dispersion relation, non-commutative field theory, Gamma Ray Bursts} %, Monte Carlo simulations}

\author{Wolfgang Bietenholz}{
  address={Instituto de Ciencias Nucleares, 
Universidad Nacional Aut\'{o}noma de M\'{e}xico, \\ 
A.P.\ 70-543, C.P.\ 04510 Distrito Federal, Mexico}
}

\begin{abstract}
We first comment on the search for a deviation
from the linear photon dispersion relation, in particular
based on cosmic photons from Gamma Ray Bursts. 
Then we consider the non-commutative space as a
theoretical concept that could lead to such a deviation, 
which would be a manifestation of Lorentz Invariance Violation. 
In particular we review a numerical study of pure $U(1)$ gauge theory
in a 4d non-commutative space. Starting from a finite
lattice, we explore the phase diagram and the extrapolation 
to the continuum and infinite volume. These simultaneous 
limits --- taken at fixed non-commutativity ---
lead to a phase of broken Poincar\'{e} symmetry, where the 
photon appears to be IR stable, despite a
negative IR divergence to one loop.
\end{abstract}

\maketitle

\section{Lorentz Invariance}

Lorentz Invariance is a central concept of relativity:
it is a global symmetry in Special Relativity, and a local
symmetry in General Relativity. Here we stay within
the framework of particle physics as described by quantum
field theory, so we refer to Special Relativity. Then this symmetry
implies that some field $\Phi$ (scalar, spinor, vector
or tensor) transforms globally in some representation $D$
of the Lorentz group $SO(1,3)$,
\begin{equation}
\Phi (x) \to U(\Lambda )^{\dagger} \Phi (x) U(\Lambda ) 
= D( \Lambda ) \Phi ( \Lambda^{-1} x) \qquad
(\Lambda \in SO(1,3) \ , \ U {\rm ~unitary}) \ .
\end{equation}

A number of high precision tests of Lorentz Invariance
involve cosmic rays, see {\it e.g.}\ Refs.\ \cite{LIVrev} 
for recent reviews. They provide perhaps the best 
chance to probe parameters not that far below the Planck scale
(in specific cases even exceeding it). Cosmic rays attain
by far the highest particle energies in the Universe (up to
$\approx 10^{20}~{\rm eV}$). In addition, the fact that they travel
over tremendous distances may be conclusive for high precision
properties even at moderate or low energies, because tiny effects
could be accumulated over a very long trajectory.
In view of the latter scenario, we discuss here 
the {\em photon dispersion relation,} as a direct 
kinematic test of Lorentz Invariance.

\section{Cosmic $\gamma$-rays}

Gamma Ray Bursts (GBRs) are emitted in powerful energy
eruptions over short periods, typically a few seconds or minutes.
Temporarily this causes the brightest spots in the sky.
They were discovered from satellites since the 70s, but nowadays
they can also be observed from ground. Their sources are small,
and there are speculations that they could be generated by the
merger of neutron stars, or of black holes. Here we are pragmatic
about their origin: in any case GRBs exist, and
we want to see what we can learn from the long journey of their 
photons through the Universe.

The distance to the source is usually well evaluated from the
redshift (if we assume the Hubble parameter $H_{0}$ to be known).
In particular, in the year 2005 a spectacular GRB
was observed \cite{swift}: it occurred at a distance of $3.9 ~{\rm Gpc}$,
{\it i.e.}\ it was emitted when the Universe was only $\approx
8 \cdot 10^{8}$ years old.
Typically the photon energies are in the range of $10^{4} \dots
10^{8} ~{\rm eV}$. Thus it is an obvious idea to use GRBs to
probe if the speed of photons is really energy independent 
\cite{Camel1}.

A systematic study of this question was performed in Ref.\ 
\cite{Ellis}. It probed the effective dispersion ansatz
\vspace*{-4.5mm}
\begin{equation} 
\vec p^{\, 2} = E^{2} \Big( 1 + \frac{E}{M} \Big) \ , \quad
v(E) =  1 - \frac{E}{M} + {\cal O} \Big( \frac{E^{2}}{M^{2}} \Big) \ ,
\end{equation}
where $M$ is a very heavy mass parameter, which might emerge
{\em somehow} from some kind of ``quantum gravity foam''.
Here we set the speed of light at photon energy $E \ll |M|$
to $c=1$. If a finite parameter $M$ exists, deviations of the group
velocity $v$ from that speed should be manifest at very high energy $E$,
{\em or} after a very long path. Ref.\ \cite{Ellis} analyzed data of
35 GRBs, detected by three satellites. In fact, photons of
higher energy tend to arrive slightly later. This could
be explained by a relative delay at the source,
$\Delta_{\rm source}$, which would then be amplified by the redshift
$z$ to the observed delay
%\begin{equation}
$ \ \Delta_{\rm obs} = \Delta_{\rm source} ( 1 + z) \, .$
%\end{equation}
So far this is standard physics. If one includes a finite parameter
$M$, this formula is modified to 
\begin{equation}
\frac{\Delta_{\rm obs}}{1+z} = \Delta_{\rm source} + a_{\rm LIV} K(z)
\ , \quad a_{\rm LIV} = \frac{\Delta E}{M H_{0}} \ ,
\end{equation}
for photon energies that differ by $\Delta E$. Here $a_{\rm LIV}$
is the Lorentz Invariance Violation (LIV) 
parameter, %$H_{0}$ the Hubble parameter,
and the function $K(z)$ is specified in Ref.\ \cite{Ellis}.
The question is now whether or not the data for 
$\Delta_{\rm obs} / (1+z)$
are compatible with a constant in $z$ (or $K(z)$). 
The authors of Ref.\ \cite{Ellis}
enhanced the errors by hand until the data became compatible with a
linear fit. % (which is something that might be discussed). 
The resulting fit is in fact almost constant, and it suggests
\begin{equation}
|M| > 1.4 \cdot 10^{25}~{\rm eV} \approx 0.001 ~ M_{\rm Planck}
\quad ({\rm with}~95~\%~{\rm C.L.}) \ .
\end{equation}
Some studies of single GRBs or blazar flares even conclude
$|M| \gsim 0.01 ~ M_{\rm Planck}$ \cite{LIVrev}. 
All these results are {\em negative} regarding the hope 
to discover new physics, but it is impressive that phenomenological
information about this energy regime is accessible at all.

Further hypothetical LIV effects, that are searched for in cosmic
photons, are a decay $\gamma \to e^{+} + e^{-}$, photon splitting, vacuum
birefringence and an irregular threshold energy for long distance
propagation through the cosmic background radiation \cite{LIVrev}.

\section{$U(1)$ gauge theory in an non-commutative space}

Here we address $U(1)$ gauge theory in an non-commutative (NC) space
as one specific theoretical framework that could lead to a 
non-linear photon dispersion relation. Due to its discontinuous
behavior in the IR limit, this approach does {\em not} belong to the 
large class of low energy effective actions --- such effective
LIV theories, including their predictions for photons, have 
also been reviewed in Refs.\ \cite{LIVrev}. Conceptual issues 
related to causality and energy positivity are discussed 
in Ref.\ \cite{Ralf}.

To describe our framework, let us start by considering a NC
Euclidean plane. The (Hermitian) position operators obey the
commutation relation
\begin{equation}
[ \hat x_{\mu} , \hat x_{\nu} ] = {\rm i} \Theta_{\mu \nu} 
= {\rm i} \theta \epsilon_{\mu \nu} \qquad (\mu , \nu = 1,2) \ ,
\end{equation}
where we assume the non-commutativity parameter $\theta$
to be constant. This implies a spatial uncertainty
relation $\Delta x_{1} \Delta x_{2} = {\cal O} (\theta )$.
In a Gedankenexperiment, Ref.\ \cite{DFR} interpreted this relation 
(in $d=4$) as the event horizon, when one tries to measure extremely short 
distances (of the order of the Planck length), which requires an enormous
energy concentration in this range.
Thus $\sqrt{ \Vert \Theta \Vert }$ could be viewed as a
``minimal length'', {\it i.e.}\ an additional constant of Nature (similar
to the parameter $1/M$ in Section 2).

A field theory on such a space is {\em non-local.} On the quantum level,
the modification of the standard geometry at short distances implies not
only UV, but also IR effects --- in particular perturbation theory
%and renormalization group concepts 
is in deep trouble due to the emergence of a new type of {\em IR 
divergences} (``UV/IR mixing'') \cite{UVIR}.

Here we consider a fully {\em non-perturbative approach},
which is even more motivated than in commutative quantum field theory.
In analogy to that case, we impose a (fuzzy) lattice structure
of spacing $a$ by means of the operator identity
\be
\exp \Big( {\rm i} \frac{2\pi}{a} \hat x_{\mu} \Big) = \hat \uno \ .
\ee
The momentum components $p_{\mu}$ commute, and they obey the usual 
periodicity over the Brillouin zone, which implies
%\bea
\be
e^{{\rm i} p_{\mu} \hat x_{\mu}} = 
e^{{\rm i} (p_{\mu} + 2 \pi/a) \hat x_{\mu}} \ \Rightarrow \  %\nn \\
\hat \uno = e^{{\rm i} (p_{\mu} + 2 \pi/a) \hat x_{\mu}} 
e^{-{\rm i} p_{\nu} \hat x_{\nu}} = \dots = \hat \uno 
\exp \Big( \frac{{\rm i}\pi}{a} \theta (p_{2} - p_{1}) \Big) \ . \nn
\ee
%\eea
At fixed $\theta$ and $a$, this means that the momenta are
{\em discrete}, and the lattice is automatically periodic.
This is of course in contrast to the commutative space.
If the lattice is $N \times N$ periodic, the momentum components
take the form 
\be
p_{\mu} = \frac{2\pi}{aN} \, n_{\mu} \quad (n_{\mu} \in \mathbb{Z}) \quad
\Rightarrow \quad \theta = \frac{1}{\pi} N a^{2} \ .
\ee
Therefore the {\em Double Scaling Limit}
\be  \label{DSL}
\{ a \to 0 \quad {\rm and} \quad N \to \infty \} \quad {\rm at} \quad
N a^{2} = {\rm const.}
\ee
leads to a {\em continuous NC plane of an infinite extent.}
We start from a finite lattice, where both the UV and the
IR sector are regular, and we simultaneously remove
both regularizations in a controlled manner.
Other procedures could easily lead to $\theta \to 0$ or
$\theta \to \infty$, which are different commutative limits. 
The requirement to link the UV and IR extrapolations is related
to the notorious UV/IR mixing (for a review, see Ref.\ \cite{Szabo}). \\

A Fourier-type analysis shows that we can return to the use of
ordinary coordinates $x_{\mu}$, if all field multiplications are turned
into {\em star products,} such as
\be  \label{starproduct}
\phi (x) \star \psi (x) := \phi (x) \exp \Big( \frac{\rm i}{2} 
\backderi \ \!\!\! _{\mu} \, \theta \, \epsilon_{\mu \nu} \! 
\forderi \ \!\!\! _{\nu} \Big) \psi (x) \ .
\ee
The $\star$-commutator 
$[ x_{\mu}, x_{\nu} ]_{\star} := x_{\mu} \star x_{\nu} - x_{\nu} \star x_{\mu}
= {\rm i} \Theta_{\mu \nu}$
makes this formulation 
plausible.

Hence the action of pure $U(1)$ gauge theory on a NC plane 
can be written as
\be
S[A] = \frac{1}{2} \int d^{2}x \, F_{\mu \nu} \star F_{\mu \nu} \ ,
\quad F_{\mu \nu} = \partial_{\mu}A_{\nu} - \partial_{\nu} A_{\mu}
+ {\rm i} g (A_{\mu} \star A_{\nu} - A_{\nu} \star A_{\mu}) \ .
\ee
This action is $\star$-gauge invariant (though this does not
hold for $F_{\mu \nu}$). Note that non-commutativity
brings in a Yang-Mills type self-interaction term.

At first sight, it looks hopeless to put this action on a lattice
and simulated it: this seems to require $\star$-unitary link
variables, $U_{x, \mu}^{\dagger} \star U_{x, \mu} = \uno$ (no sum),
which can only be constructed and updated with stringent 
constraints given by the complete configuration.
There is a way out, however, namely the mapping onto a 
{\em matrix model}.

A long time ago, and in a different context, Gonz\'{a}lez-Arroyo
and Okawa introduced the {\em twisted Eguchi-Kawai (TEK) model}
\cite{GAO}. It is formulated on a single point, with the action 
\be
S_{\rm TEK}[U] = N \beta \sum_{\mu \neq \nu} Z_{\mu \nu}
{\rm Tr} [ U_{\nu}^{\dagger} U_{\mu}^{\dagger} U_{\nu} U_{\mu}] \qquad
( \beta = 1/ g^{2} ) \ ,
\ee
where $U_{\mu}$ ($\mu = 1 , 2$ in $d=2$) are unitary $N \times N$
matrices (which emerged from link variables after dimensional 
reduction of $U(N)$ lattice gauge theory),
and $Z_{21} = Z_{12}^{*} = \exp ( 2 \pi {\rm i} n /N)$ is the twist
factor ($n \in \mathbb{Z}$, $n~{\rm mod}~ N \neq 0$). 
Much later it turned out that this
model is equivalent to NC $U(1)$ gauge theory on a %(periodic)
$N \times N$ lattice: in particular the algebras were shown to be
identical (Morita equivalence) \cite{Morita}. 

Referring to the historic interpretation of the matrices as
contracted link variables, it is obvious to formulate 
(rectangular) Wilson loops in the TEK model as
\be
W_{\mu \nu} (I \times J) = \frac{1}{N} \, Z_{\mu \nu}^{I \cdot J} \,
{\rm Tr} [ U_{\nu}^{\dagger ~ J} U_{\mu}^{\dagger ~ I} 
U_{\nu}^{J} U_{\mu}^{I} ] \ .
\ee
Mapping this term back to the NC gauge theory leads to a sensible
definition of a NC Wilson loop \cite{Wloop}. (There are no ``closed 
loops'' on a NC plane, but the essential criterion is
$\star$-gauge invariance).
A specific Wilson loop is complex in general,
but the property $W_{\mu \nu} = W_{\nu \mu}^{*}$ guarantees that
the action is real positive, since it involves a sum over both
plaquette orientations. 
Therefore the TEK model formulation is suitable for 
Monte Carlo simulations, {\it i.e.} for a non-perturbative
study of NC gauge theory.

\section{The fate of the non-commutative photon}

Let us now address the question how a possible non-commutativity
could deform the photon dispersion relation. A 1-loop result of the form 
\be  \label{1loop}
E^{2} = \vec p^{\, 2} + \frac{C g^{2}}{(p \Theta)^{2}} \qquad
(C = {\rm const.})
\ee
was first derived in Ref.\ \cite{MST}. Hence one could even be
tempted to deduce a {\em lower} bound on $\Vert \Theta \Vert$
from the observed high-precision linear dispersion relation
({\em if} $\Theta \neq {\bf 0}$)
\cite{CamelNC}. However, further calculations revealed that the
constant $C$ is actually {\em negative}, $C = -2/\pi^{2}$ \cite{LLT}. 
This apparent IR instability is worrisome indeed: it seemed questionable if
NC QED does have a ground state, {\it i.e.}\ if it actually represents a
well-defined quantum theory. Hence people quickly proceeded to a 
supersymmetric version of this model, where the negative 
IR divergence is cancelled \cite{LLT}.
(It is not visible either, of course, when one treats the model
classically \cite{Rivelles}.)
%\footnote{Of course, this 
%problem does not occur in a classical treatment of the NC photon
%\cite{Rivelles}.}

However, we were not happy with this combination of daring hypotheses,
so we did not assume supersymmetry in addition.
% (anyhow there is some excess of supersymmetry in this workshop).
We wanted to verify if the NC photon as such is really
ill-defined \cite{NCQED}. To this end, we considered 
a ``minimally'' NC Euclidean space-time, which involves a NC plane,
$[ \hat x_{1}, \hat x_{2} ] = {\rm i} \theta = {\rm const.}$, plus
a commutative plane $(x_{3},t)$ which includes the Euclidean time $t$.
(A NC time coordinate would break reflection positivity, so
in that case the transition to the Minkowski signature would 
no be on safe grounds \cite{OSax}. Commutative time also alleviates
the problems with causality.)

\begin{figure}[h!]
\hspace*{1mm}
\includegraphics[height=.24\textheight,angle=270]{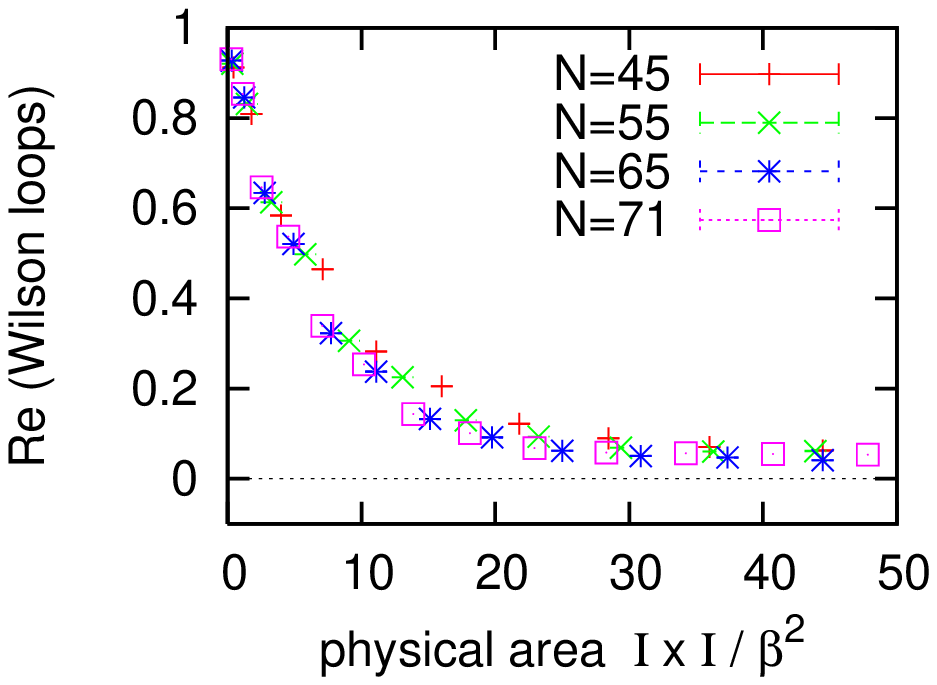}
\hspace*{-5mm}
\includegraphics[height=.24\textheight,angle=270]{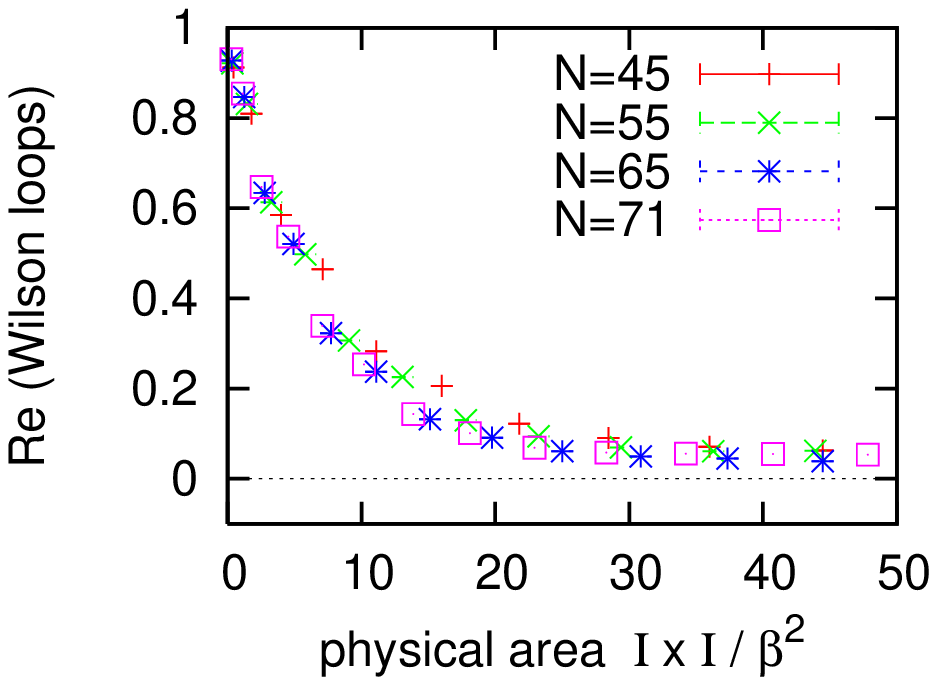}
\hspace*{-5mm}
\includegraphics[height=.24\textheight,angle=270]{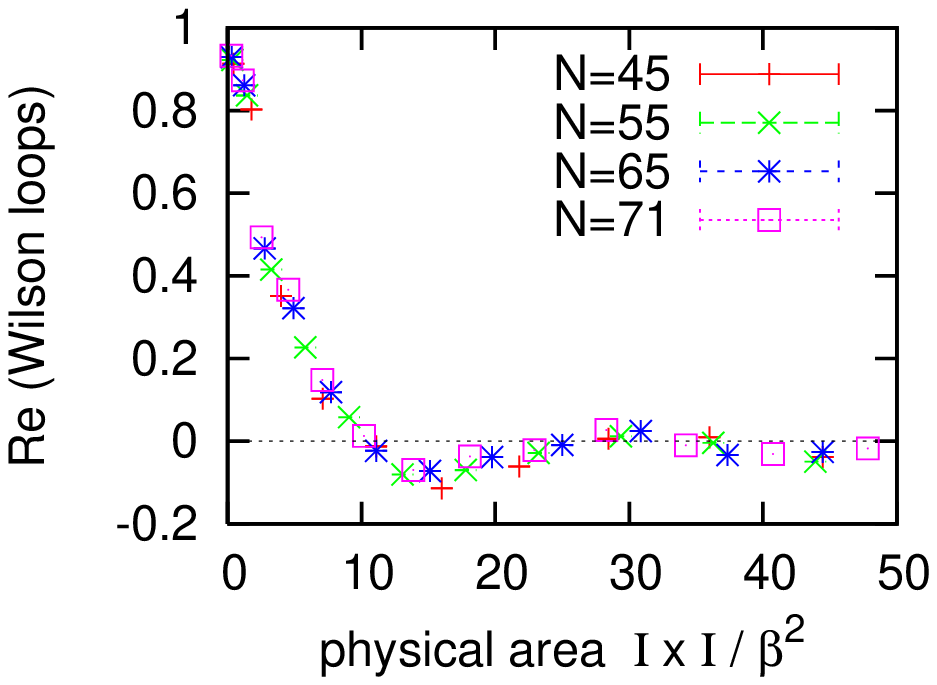}
\caption{The real part of square-shaped Wilson loops at a constant
ratio $N / \beta^{2}= 20$. We show results in the commutative
plane (left), a mixed plane (center) and the NC plane (right).
In the former two planes the Wilson loops are real due to the
symmetry in the signs of $x_{3}$ and $t$. In the NC plane
$\langle W_{12} \rangle$ 
has a complex phase which rises linearly in the area
for large loops. (This Aharonov-Bohm-type effect was also observed
in 2d NC $U(1)$ gauge theory \cite{2dNCU1}, though with some
shape dependence of the Wilson loop expectation values \cite{BBT}.)
In all cases we observe an excellent Double Scaling behavior.}
\label{DSLillu}
\end{figure}

For the commutative plane we used the standard regularization
of an $L \times L$ lattice, while the NC plane was treated with
a TEK model as described in Section 3, {\it i.e.}\ we put a $N \times N$
matrix model on each of the $L^{2}$ lattice sites ($L \approx N$).
The first challenge for the numerical measurements was the identification
of a {\em scale,} in order to define the Double Scaling Limit (\ref{DSL}).
A simple ansatz for the lattice spacing, $a \propto 1 / \beta$,
was successful. It corresponds to a Double Scaling Limit which
keeps the ratio $N / \beta^{2}$ constant. Fig.\ \ref{DSLillu} 
illustrates, for the example $N / \beta^{2} = 20$, that the Wilson 
loops $\langle W_{\mu \nu} \rangle$
of different sizes, in various planes, are in fact stable as we
vary $N$ and $\beta$ accordingly.

\begin{figure}[h!]
\includegraphics[height=.335\textheight,angle=270]{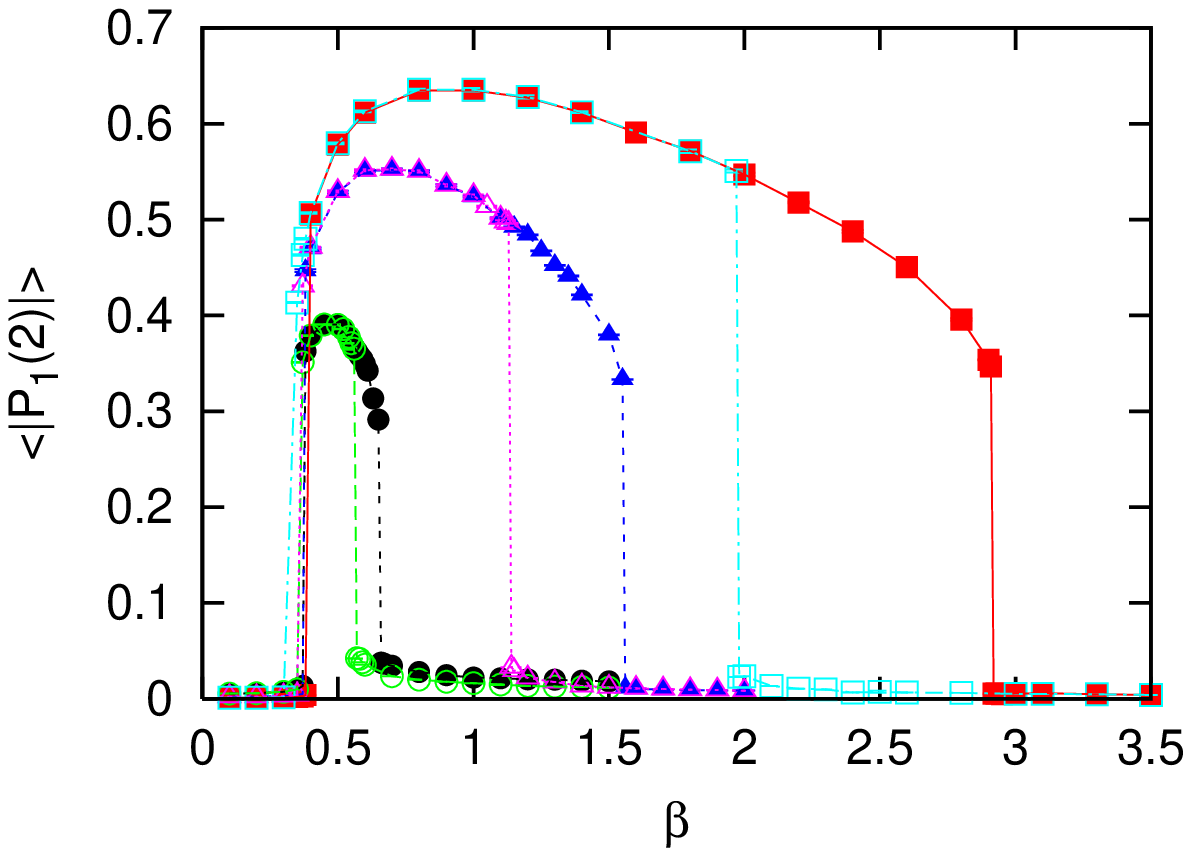}
\includegraphics[height=.335\textheight,angle=270]{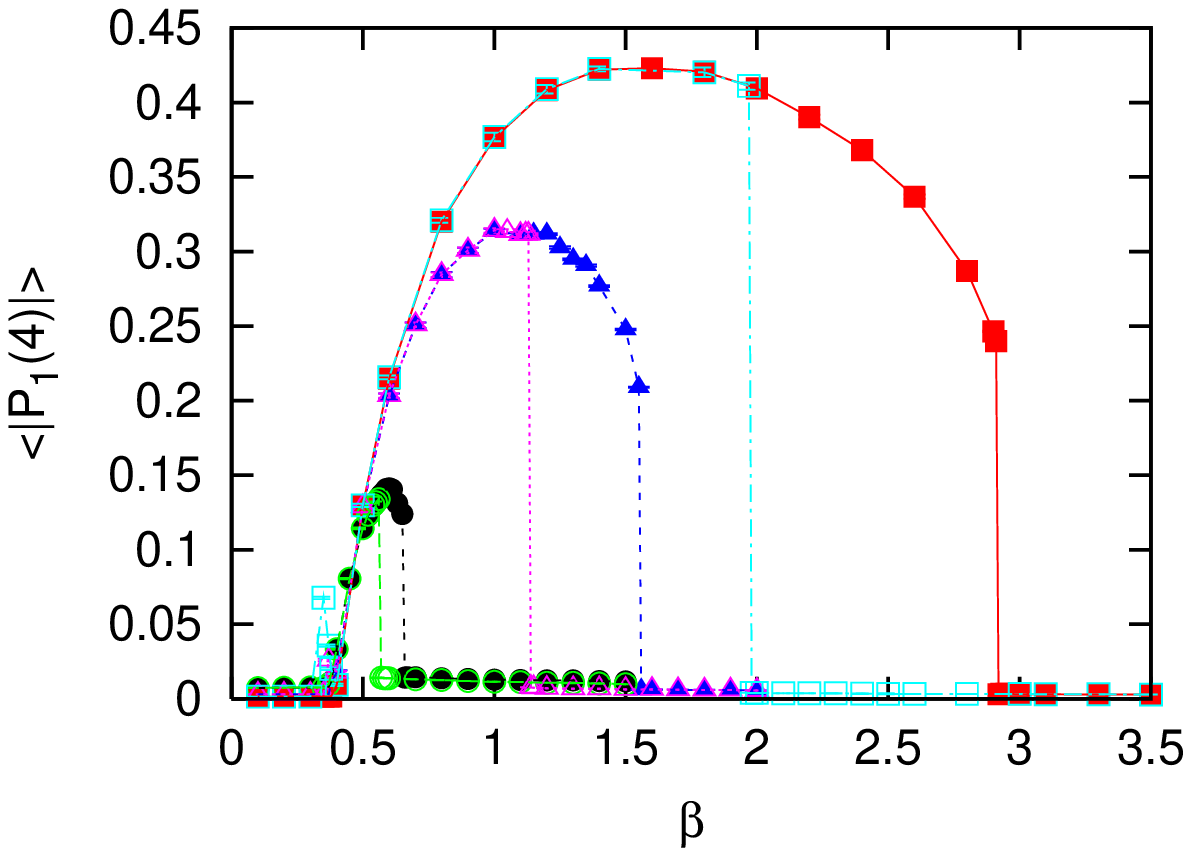}
\caption{The measured vacuum expectation values for the open Wilson
lines of length 2 and 4 lattice units, at $N = 15,\, 25, \, 35$
(from left to right), and $\beta$ fixed by the Double Scaling condition. 
We observe symmetric phases at strong and at weak coupling, but at moderate
coupling we discovered a phase of spontaneously broken translation
symmetry. The weak coupling transition has a clear hysteresis
behavior.}
%, which implies that it is of first order.}
\label{poly24}
\end{figure}

Next we measured the open Wilson lines $P_{\mu}(n)$ in $\mu$-direction
of length $n$ in lattice unit. In dimensional units this line follows the
vector
\be  \label{openW}
\Theta_{\mu \nu} p_{\nu} = n a \hat \mu \qquad (\hat \mu ~:~{\rm unit~
vector~in~}\mu ~{\rm direction}) \ .
\ee
This observable is $\star$-gauge invariant, in agreement with the
non-locality of the $\star$-gauge transformation \cite{Wloop}. Such an open 
Wilson line carries momentum $p$ given in eq.\ (\ref{openW}). It is 
therefore a suitable order parameter to detect if translation invariance 
in the NC plane is broken.
Fig.\ \ref{poly24} shows a symmetric behavior at strong coupling
($\beta \lsim 0.35$) and again at weak coupling, where the transition value
of $\beta$ depends on $N$. This agrees with the expectation based on
strong and weak coupling expansions, but in between we discovered a 
{\em phase of broken symmetry,} 
which was not expected from any analytical consideration.
For the weak coupling transition, Fig.\ \ref{poly24} shows a marked
hysteresis (the two curves for fixed $N$ refer to increasing and decreasing
coupling strength), which is characteristic for a first order phase transition.
\vspace*{-2mm}
\vspace*{-1mm}

\begin{figure}[h!]
\includegraphics[height=.33\textheight,angle=0]{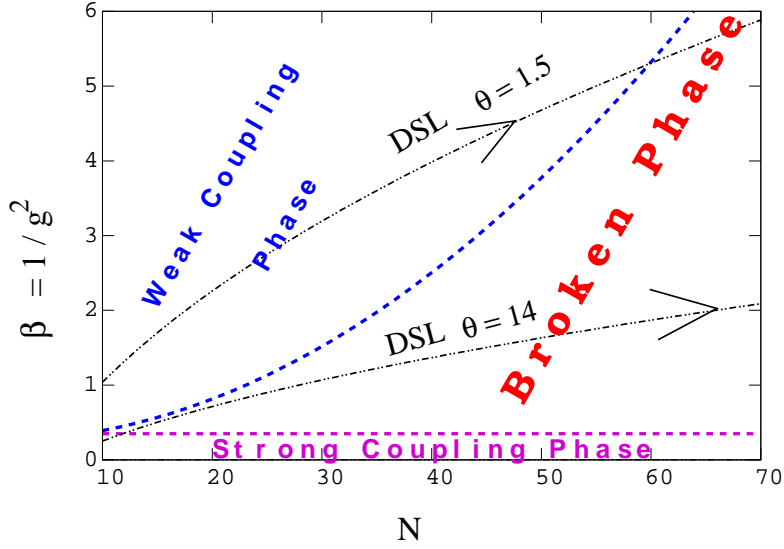}
\caption{The phase diagram for pure $U(1)$ gauge theory in 4 dimensions,
involving a (spatial) NC plane. At fixed size $N$,
Poincar\'{e} symmetry is intact at strong
and at weak coupling, but it is spontaneously broken in a moderate
coupling phase. The Double Scaling Limits (DSL) --- taken at fixed 
non-commutativity parameter $\theta$ --- 
lead to the {\em broken} phase, which is therefore relevant 
for the NC continuum limit. In that phase, the (negative) IR singularity
is not observed, and the system could be sensible. \vspace*{-6mm}}
\label{phasedia}
\end{figure}

Fig.\ \ref{phasedia} shows the corresponding phase diagram.
For any $N$ the strong coupling transition is at $\beta \approx 0.35$
(cf.\ Fig.\ \ref{poly24}), whereas the weak coupling transition
occurs at a value of $\beta \propto N^{2}$. This means that
the Double Scaling Limits --- at fixed $\theta \propto N a^{2} \propto
N / \beta^{2}$ --- all lead to the {\em broken phase,} and not to the weak
coupling phase as one might have expected. Hence it is the broken phase
which is really relevant for the question if observables in NC $U(1)$
gauge theory are stable or not. The 1-loop result (\ref{1loop}) refers
to the weak coupling phase; it is therefore not relevant for the
NC continuum limit (the Double Scaling Limit).
In the broken phase we found stability for all 
observables that we measured \cite{NCQED}.
This holds in particular for our explicit results for the
dispersion relation of the NC photon. It was measured from the
exponential decay of the Wilson line correlation function, separated
in Euclidean time. We set the momentum components
in the NC plane to zero, hence we measured the energy
$E(p = p_{3})\vert_{p_{1}=p_{2} = 0}\, $.

\begin{figure}[h!]
\includegraphics[height=.33\textheight,angle=270]{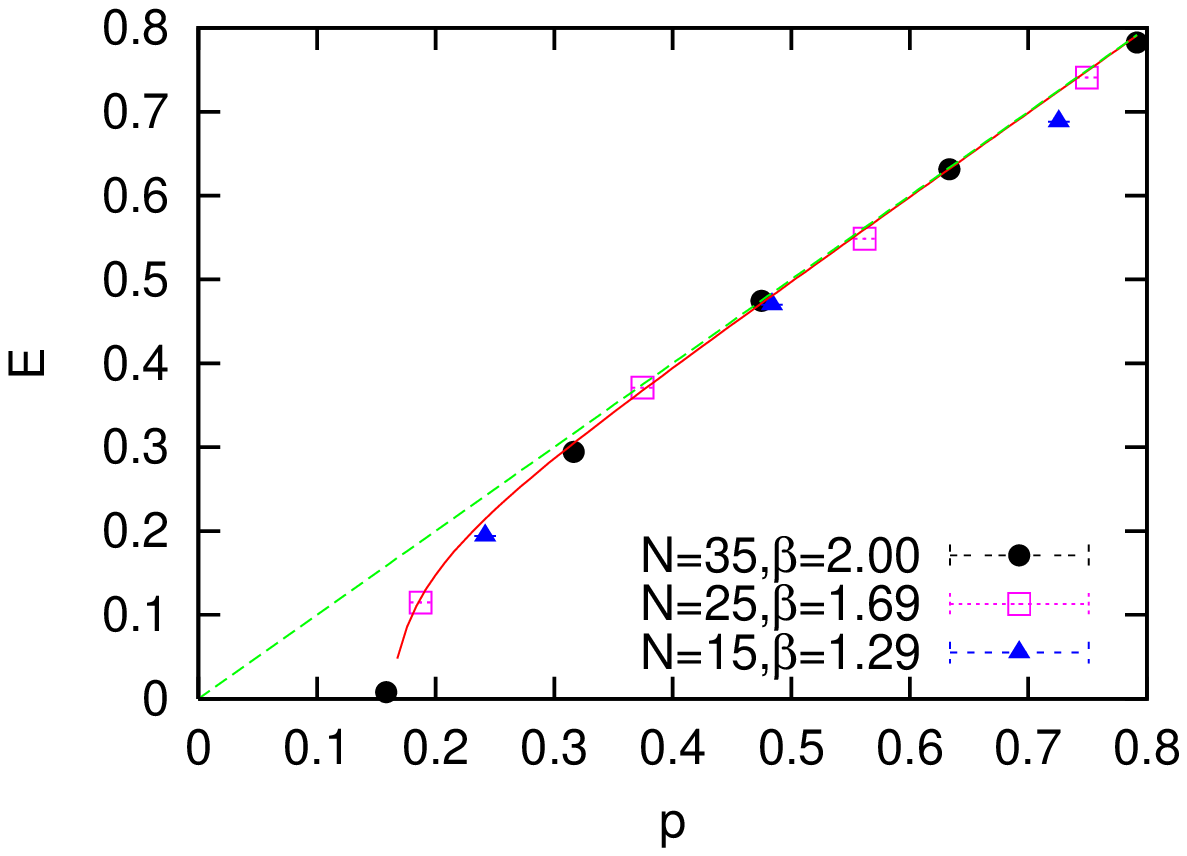}
\includegraphics[height=.33\textheight,angle=270]{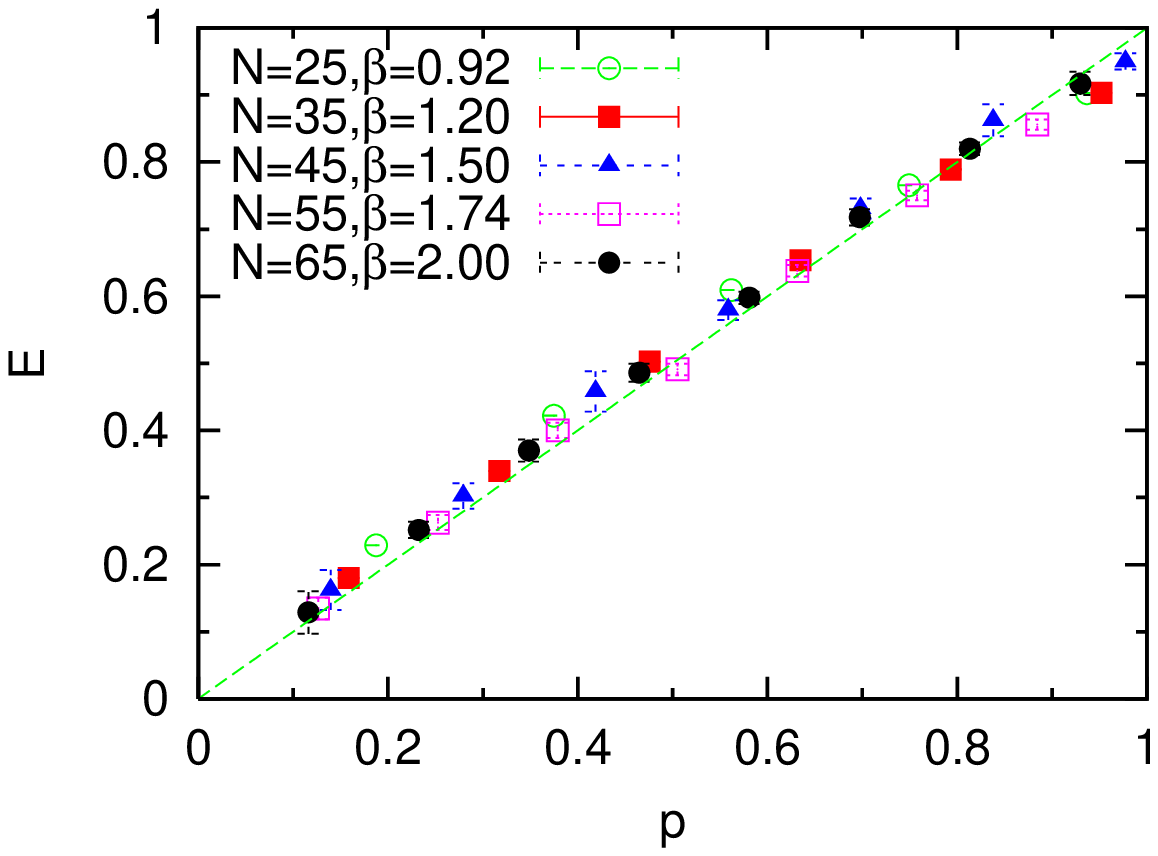}
\caption{The dispersion relation of the NC photon in the symmetric
weak coupling phase (on the left), and in the phase of broken Poincar\'{e}
symmetry (on the right). We show the energy as a function of the commutative
momentum component $p_{3}$, $E(p = p_{3})\vert_{p_{1}=p_{2} = 0}\, $.
%at vanishing momentum
%components in the NC plane. 
Our result in the symmetric phase is consistent with the IR 
instability, which was predicted perturbatively \cite{LLT}.
(In contrast, a positive IR divergence has been observed in the
NC $\lambda \phi^{4}$ model, where translation symmetry breaks as well 
\cite{NCphi4}.)
The physically relevant result, however, is the dispersion in the broken
phase, where the Double Scaling Limit ends up, and there we observe an
{\em IR stable} behavior.}
\label{disprel}
\end{figure}

Fig.\ \ref{disprel} (on the left) refers to
the (symmetric) weak coupling phase. We see an energy dip at low momenta.
This feature is fully consistent with the prediction (\ref{1loop}) based on 
perturbation theory, which some people denote as ``tachyonic behavior''.
(Of course we cannot measure negative energies based on
the decay of 2-point functions, but the feature observed
%on the left-hand-side of Fig.\ \ref{disprel} 
here matches exactly such an IR instability).

However, as we pointed out above, the phase which really matters
for a continuum limit to a NC space of infinite extent is the
phase of broken Poincar\'{e} symmetry. The result in that
phase is illustrated on the right-hand-side of Fig.\ \ref{disprel}.
It follows the standard linear dispersion relation. Hence we do not
see any non-commutativity effect for the dispersion in the commutative
plane, but we do see that the disastrous negative IR divergence 
disappears. 
Our attempts to measure the dispersion relation also at finite
momentum components in the NC plane were not successful, since we
did not obtain an exponential decay of the Wilson line correlation 
functions. 
%It is conceivable that this correlator has indeed an
%irregular behavior, so that an energy cannot be determined in the
%usual way, or that it takes much larger systems for an exponential
%decay to set in.

\vspace*{-18mm}

\section{Conclusions}

Lorentz Invariance Violation (LIV) has been searched for intensively
in recent years, but so far it has not been observed anywhere.
We reviewed a failed attempt which checked the photon dispersion relation
based on Gamma Rays Bursts.

As one of the theoretical frameworks which could give rise to LIV
we discussed quantum field theory --- in particular
pure $U(1)$ gauge theory --- in a non-commutative space.
Our non-perturbative study \cite{NCQED} was based on Monte Carlo 
simulations in a space of Euclidean signature with one (spatial) 
non-commutative plane.
The Double Scaling Limit extrapolates simultaneously to the
continuum and to infinite volume at a fixed non-commutativity 
parameter. It leads to a phase of spontaneously broken Poincar\'{e} 
symmetry, which corresponds to a moderate coupling strength
in finite volume. 

In this limit we found evidence for convergent
observables. This suggests that this model might be
renormalizable, despite severe problems in its perturbative 
expansion \cite{LLT}. In particular the photon dispersion in the
Double Scaling Limit appears IR stable, though we could only
measure it in the commutative plane. If one measures
a deviation from the linear behavior in an NC direction, the
confrontation with GRB data could establish a robust bound
on $\Vert \Theta \Vert$ (the norm of the non-commutativity 
tensor) in Nature. So far we provided evidence that the photon
may survive in a NC space, although perturbation theory found
an negative IR divergence (that result only holds in the
weak coupling phase, which turned out to be irrelevant for 
the NC continuum limit).

As an outlook on the phenomenological side, we mention that
the Fermi Gamma-ray Space Telescope has been in orbit since
June 2008. It monitors hundreds of GRBs 
per year, with sensitivity to photon energies
$E \simeq 8 ~ {\rm keV} \dots 300 ~ {\rm GeV}$ \cite{Lamon}. 
This could further boost the precision of Lorentz Invariance 
tests on cosmic photons \cite{Fermi}, cf.\ Section 2. \\ 

%\begin{theacknowledgments}
\vspace*{-2mm}
\noindent
{\small{\bf Acknowledgment: } 
I thank Jun Nishimura, Yoshiaki Susaki and Jan Volkholz for their
collaboration in the work that I summarized here,
and Frank Hofheinz for his assistance.}
%I also thank Frank Hofheinz for his valuable assistance in this project.}
\vspace*{-2.5mm}
%\end{theacknowledgments}

\bibliographystyle{aipproc}   % if natbib is available

\begin{thebibliography}{10}
\vspace*{-1mm}

\bibitem{LIVrev} D.\ Mattingly, {\em Living Rev.\ Rel.}\ {\bf 8}, 5 (2005).
T.\ Jacobson, S.\ Liberati and D.\ Mattingly,
% Lorentz violation at high energy: 
% Concepts, phenomena and astrophysical constraints.
{\em Annals Phys.}\ {\bf 321}, 150--196 (2006). % astro-ph/0505267
W.\ Bietenholz,
% Cosmic Rays and the Search for a Lorentz Invariance Violation.
arXiv:0806.3713 [hep-ph];
{\em Fortschr.\ Phys.}\ {\bf 57}, 505--513 (2009). %arXiv:0812.3564 [hep-th]
M.\ Galaverni and G.\ Sigl, {\em Phys.\ Rev.}\ {\bf D 78}, 063003 (2008).
S.\ Liberati and L.\ Maccione,
{\em Ann.\ Rev.\ Nucl.\ Part.\ Sci.}\ {\bf 59}, 245--267 (2009).
L.\ Shao and B.-Q.\ Ma, arXiv:1007.2269 [hep-ph].

\bibitem{swift} http://www.nasa.gov/mission$\_$pages/swift/bursts/farthest$\_$grb.html

\bibitem{Camel1} G.\ Amelino-Camelia {\it et al.}, 
%J.\ Ellis, N.E.\ Mavromatos, D.V.\ Nanopoulos and S.\ Sarkar,
{\em Nature} {\bf 393}, 763--765 (1998).

\bibitem{Ellis} J.R.\ Ellis {\it et al.}, 
%N.E.\ Mavromatos, D.V.\ Nanopoulos, A.S.\ Sakharov and E.K.G.\ Sarkisyan,
% Robust limits on Lorentz violation from gamma-ray bursts
{\em Astropart.\ Phys.}\ {\bf 25}, 402--411 (2006); % astro-ph/0510172
erratum, arXiv:0712.2781 [astro-ph].

\bibitem{Ralf} V.A.\ Kosteleck\'{y} and R.\ Lehnert,
% Stability, causality, and Lorentz and CPT violation
{\em Phys.\ Rev.}\ {\bf D 63}, 065008 (2001).
% hep-th/0012060

\bibitem{DFR} S.\ Doplicher, K.\ Fredenhagen and J.E.\ Roberts,
%The Quantum structure of space-time at the Planck scale and quantum fields.
{\em Commun.\ Math.\ Phys.}\ {\bf 172}, 187--220 (1995). % hep-th/0303037

\bibitem{UVIR} S.~Minwalla, M.~Van Raamsdonk and N.~Seiberg,
{\em JHEP} \textbf{02}, 020 (2000).

\bibitem{Szabo} R.J.\ Szabo,
% Quantum field theory on noncommutative spaces.
{\em Phys.\ Rept.}\ {\bf 378}, 207--299 (2003). % hep-th/0109162

\bibitem{GAO} A.\ Gonz\'{a}lez-Arroyo and M.\ Okawa,
% The Twisted Eguchi-Kawai Model: A Reduced Model 
% for Large N Lattice Gauge Theory.
{\em Phys.\ Rev.}\ {\bf D 27}, 2397--2411 (1983).

\bibitem{Morita} H.~Aoki, N.~Ishibashi, S.~Iso, H.~Kawai, 
Y.~Kitazawa and T.~Tada, 
{\em Nucl.\ Phys.}\ {\bf B 565}, 176--192 (2000).
J.~Ambj{\o}rn, Y.~Makeenko, J.~Nishimura and R.J.~Szabo, 
{\em JHEP} {\bf 05}, 023 (2000).

\bibitem{Wloop} N.~Ishibashi, S.~Iso, H.~Kawai and Y.~Kitazawa,
{\it Nucl.\ Phys.} \textbf{B 573}, 573--593 (2000).
D.J.~Gross, A.~Hashimoto and N.~Itzhaki,
%``Observables of non-commutative gauge theories'',
{\em Adv.\ Theor.\ Math.\ Phys.}  {\bf 4}, 893--928 (2000).
%{\tt [hep-th/0008075]};

\bibitem{MST} A.\ Matusis, L.\ Susskind and N.\ Toumbas, 
% The IR/UV Connection in the Non-Commutative Gauge Theories
{\em JHEP} {\bf 0012}, 002 (2000). % [hep-th/0002075]

\bibitem{CamelNC} G.\ Amelino-Camelia, L.\ Doplicher, S.-K.\ Nam
and Y.-S.\ Seo,
% % Phenomenology of particle production and propagation in 
% % string motivated canonical noncommutative space-time
{\em Phys.\ Rev.}\ {\bf D 67}, 085008 (2003). % hep-th/0109191

\bibitem{LLT} F.\ Ruiz Ruiz, 
% Gauge-fixing independence of IR divergences in non-commutative 
% U(1), perturbative tachyonic instabilities and supersymmetry
{\em Phys.\ Lett.}\ {\bf B 502}, 274--278 (2001). % hep-th/0012171
K.\ Landsteiner, E.\ Lopez and M.H.G.\ Tytgat,
% Instability of noncommutative SYM theories at finite temperature
{\em JHEP} {\bf 0106}, 055 (2001). % hep-th/0104133 
M.\ Van Raamsdonk,
% The Meaning of infrared singularities in noncommutative gauge theories.
{\em JHEP} {\bf 0111}, 006 (2001). % hep-th/0110093

\bibitem{Rivelles} T.\ Mariz, J.R.\ Nascimento and V.O.\ Rivelles,
% Dispersion Relations in Noncommutative Theories
{\em Phys.\ Rev.}\ {\bf D 75}, 025020 (2007).
% hep-th/0609132

\bibitem{NCQED} W.\ Bietenholz, J.\ Nishimura, Y.\ Susaki and J.\ Volkholz,
% A Non-perturbative study of 4-D U(1) non-commutative gauge theory: 
% The Fate of one-loop instability
{\em JHEP} {\bf 0610}, 042 (2006). % hep-th/0608072

\bibitem{OSax} K.\ Osterwalder and R.\ Schrader,
{\em Commun.\ Math.\ Phys.} {\bf 31}, 83--112 (1973).

\bibitem{2dNCU1} W.\ Bietenholz, F.\ Hofheinz and J.\ Nishimura,
% A Non-Perturbative Study of Gauge Theory on a Non-Commutative Plane
{\em JHEP} {\bf 0209}, 009 (2002). % hep-th/0203151

\bibitem{BBT} W.\ Bietenholz, A.\ Bigarini and A.\ Torrielli,
% Area-preserving diffeomorphisms in gauge theory 
% on a non-commutative plane: A Lattice study.
{\em JHEP} {\bf 0708}, 041 (2007). % arXiv:0705.3536 [hep-lat]

\bibitem{NCphi4} W.\ Bietenholz, F.\ Hofheinz and J.\ Nishimura,
% Phase diagram and dispersion relation of the noncommutative 
% lambda phi**4 model in d = 3
{\em JHEP} {\bf 0406}, 042 (2004). % hep-th/0404020
 
\bibitem{Lamon} R.\ Lamon,
% GLAST and Lorentz violation
{\em JCAP} {\bf 0808}, 022 (2008) 022. % arXiv:0805.1219 [astro-ph]

\bibitem{Fermi} Fermi GBM/LAT Collaborations (A.A.\ Abdo {\it et al.}),
% A limit on the variation of the speed of light arising 
% from quantum gravity effects
{\em Nature} {\bf 462}, 331--334 (2009). %arXiv:0908.1832 [astro-ph.HE].
L.\ Shao, Z.\ Xiao and B.-Q.\ Ma,
% Lorentz violation from cosmological objects with 
% very high energy photon emissions
{\em Astropart.\ Phys.}\ {\bf 33}, 312--315 (2010).
% arXiv:0911.2276 [hep-ph] 

\end{thebibliography}

\end{document}